\documentclass[12pt,thmsa]{article}
\usepackage{amsfonts}

\input{tcilatex}
\begin{document}

\author{Karl-Georg Schlesinger \qquad \\
%EndAName
Erwin Schr\"{o}dinger Institute for Mathematical Physics\\
Boltzmanngasse 9\\
A-1090 Vienna, Austria\\
e-mail: kgschles@esi.ac.at}
\title{A universal symmetry structure in open string theory }
\date{}
\maketitle

\begin{abstract}
In this paper, we arrive from different starting points at the conclusion
that the symmetry given by an action of the Grothendieck-Teichm\"{u}ller
group $GT$ on the so called extended moduli space of string theory can not
be physical - in the sense that it does not survive the inclusion of general
nonperturbative vacua given by boundary condition on the level of two
dimensional conformal field theory - but has to be extended to a quantum
symmetry given by a self-dual, noncommutative, and noncocommutative Hopf
algebra $\mathcal{H}_{GT}$. First, we show that a class of two dimensional
boundary conformal field theories always uniquely defines a trialgebra and
find $\mathcal{H}_{GT}$ as the universal symmetry of such trialgebras (in
analogy to the definition of $GT$ as the universal symmetry of
quasi-triangular quasi-Hopf algebras). Second, we argue in a more heuristic
approach that the $\mathcal{H}_{GT}$ symmetry can also be found in a more
geometric picture using the language of gerbes.
\end{abstract}

\section{Introduction}

For the case of perturbative vacua in string theory, i.e. two dimensional
superconformal field theory, there is a large body of results on certain
symmetry structures which are linked to these. First of all, there are a
number of reconstruction theorems (see \cite{CP}, \cite{FK}) which allow to
view the fusion structure of these superconformal theories as given by the
representation category of a quasitriangular quasi-Hopf algebra. Beyond
this, one can find a universal symmetry of all quasitriangular quasi-Hopf
algebras in the form of the so called Grothendieck-Teichm\"{u}ller group $GT$
(see \cite{Dri}). The basic idea is the following: Ask for the possibility
to transform the $R$-matrix and the Drinfeld associator $\alpha $ of a
quasi-triangular quasi-Hopf algebra $H$ while keeping the rest of the
structure of $H$ completely fixed. In order to define a nontrivial group
from these transformations, one has to take a certain closure by including
formal deformations of $R$ and $\alpha $ in the sense of a class of formal
power series. Doing this, one arrives at $GT$ (for the full technical
details of the definition of $GT$ we refer the reader to \cite{Dri}, for a
convenient short description see also \cite{CP}). As a consequence of these
two levels of symmetry structure - the Hopf algebra structures of two
dimensional superconformal field theory and the universal symmetry of $GT$ -
there exists an action of $GT$ on the so called extended moduli space of two
dimensional superconformal field theories (see \cite{Kon 1994}, \cite{Wit}
for the introduction and general structure of this space and \cite{Kon 1999}%
, \cite{KoSo} for the $GT$-action on it). Mathematically, this space is
expected to be describable as the moduli space of a triangulated version of $%
A_\infty $-categories (see \cite{Kon 1994}). On the one hand, this extended
moduli space has been related to $D$-branes, i.e. to boundary conditions for
open strings (see e.g. \cite{GZ}, \cite{Laz}). On the other hand, even this
extended moduli space (it is extended in comparison to the usual moduli
space of a two dimensional superconformal field theory which is locally
generated by the truely marginal operators) is restricted to see non-metric
degrees of freedom, only (see \cite{Kon 1994}, \cite{Wit}), i.e. it
basically describes a topological string theory.

An understanding of the structure of the relevant moduli space of vacua for
string theory is considered to be of outermost relevance for the long term
goal of a full fledged nonperturbative and background independent
formulation of the theory. Finding a symmetry like the $GT$ action on
extended moduli space is of deep principal interest. The decisive question
is, then, if this symmetry is of physical relevance. If it would be, one
could hope that it tells an important lesson about the ultimate nature of a
complete formulation of string theory. After all, we know from nearly all of
the examples of theories in physics that these are to a high degree
determined by knowing the fundamental symmetry principle. The examples range
from the Galilei invariance of classical mechanics and the Lorentz
invariance of the Maxwell equations up to the gauge symmetries in elementary
particle physics and the diffeomorphism invariance of general relativity.

A first step toward the question of a physical relevance of the $GT$%
-symmetry is taken by asking the following more concrete question, instead:
Does the $GT$-action extend to a physically relevant full moduli space of
open string theory, i.e. to general boundary conditions in the
non-topological setting (formulated more concisely: to a general moduli
space of two dimensional boundary conformal field theories)? We will
sometimes call this space simply the full moduli space, in the sequel.

We will in this paper show that the $GT$-symmetry can not extend to the full
moduli space, i.e. in this sense the $GT$-symmetry can not be physically
relevant (one can not take the $GT$-invariance as a candidate for a physical
principle in full non-topological string theory). But we find that there is
a self-dual, noncommutative, and noncocommutative Hopf algebra $\mathcal{H}%
_{GT}$ which extends the $GT$-symmetry to the full moduli space. In more
physical terms, we can view this as saying that the $GT$-symmetry does not
survive the inclusion of general nonperturbative vacua but that in the full
setting a quantum analog of this classical symmetry is seen.

We will first arrive at this result by starting from the question of what
takes the role of the Hopf algebra symmetries behind two dimensional
superconformal field theories if we pass to the setting of boundary
conformal field theories. We then ask for the universal symmetry of these
algebraic structures, replacing the $GT$-symmetry of quasi-triangular
quasi-Hopf algebras. This will be the content of section 2.

In section 3, we approach the same problem from a more geometrical
perspective: Boundary conditions in the two dimensional conformal field
theories geometrically are linked to the appearance of a 2-form field $B$
with 3-form field strength $H$. Compared to closed 2-forms (which define
connections on vector bundles), closed 3-forms - like $H$ - are linked to
connections on gerbes which, roughly speaking, can be seen as principal
bundles with categories as their fibers. Paralleling the way in which vector
bundles define by Serre duality finite-dimensional projective modules over
the algebra $\mathcal{C}^\infty \left( M\right) $ of smooth functions on the
base manifold $M$, a class of gerbes induces $\mathcal{C}^\infty \left(
M\right) $-linear ring categories. We will see in a simple but instructive
example of such a category $\mathcal{C}$ that the analog of the Hochschild
cohomology of a projective module for $\mathcal{C}$ is, again, related to an
action of the Hopf algebra $\mathcal{H}_{GT}$. So, we can argue that the
more geometrical picture of gerbes leads to the same conclusion concerning a
universal $\mathcal{H}_{GT}$-symmetry. Finally, we discuss the conjecture
that the $\mathcal{H}_{GT}$-symmetry should not only be stable against
quantization but that in this case the classical and the quantum description
should even be equivalent.

Section 4 contains some concluding remarks.

\bigskip

\section{Trialgebras and boundary conformal field theory}

Two well known results are the starting point for the considerations in this
section: First, starting from a three dimensional topological quantum field
theory, one can get a two dimensional conformal field theory as living on
the boundary of the 3-manifold (\cite{FFFS}, \cite{Wit 1989}). By the fact
that the boundary of a boundary is empty, one can get only conformal field
theories without boundary in this way. Second, three dimensional topological
field theories can be formulated in a purely algebraic way as vector space
valued functors on the category of three dimensional cobordisms (\cite{Ati})
and these functors can in turn be shown to be constructible from modular
categories (see \cite{Tur}). So, two dimensional conformal field theories
can be constructed starting from certain modular categories (see \cite{Seg}
for a different approach leading to this result). In \cite{KL} an algebraic
framework is presented motivated by the wish to extend this approach to the
case of two dimensional boundary conformal field theories. The algebraic
notion of topological quantum field theory of \cite{Ati} is extended
(algebraically by using double functors, i.e. morphisms between double
categories) to include 3-manifolds which do not only have boundaries but
where also corners on the boundaries are allowed. This introduces the
necessary freedom for two dimensional boundary conformal field theories to
appear on the boundary of the 3-manifold (see \cite{KL} for the details).
The central result of \cite{KL} is the following:

The extended topological quantum field theories in the above sense are in
one to one correspondence with $\Bbb{C}$-linear, bounded (i.e. equivalent to
a category of finite dimensional modules over a finite dimensional algebra),
balanced, abelian, rigid, braided, monoidal categories $\mathcal{C}$
together with a self-dual Hopf algebra object $H$ together with a special
self-duality morphism in them (see, again, \cite{KL} for the detailed
definitions).

So, the algebraic framework introduced in \cite{KL} leads to the conclusion
that at least a class of two dimensional boundary conformal field theories
can be defined starting from such categories $\mathcal{C}$ together with a
Hopf algebra object $H$. It is shown in \cite{KL} that the input data of 
\cite{Tur} give a special case of the data $\left( \mathcal{C},H\right) $.
This means that the data, given by the introduction of a Hopf algebra object 
$H$, are directly linked to boundary conditions for the conformal field
theory. We will, in the sequel, always assume that one can make use of the
topological approach of \cite{KL} in studying two dimensional boundary
conformal field theory.

\bigskip

In order to proceed, we now introduce a new algebraic concept, called a
trialgebra:

\bigskip

\begin{definition}
A trialgebra $(A,*,\Delta ,\cdot )$ with $*$ and $\cdot $ associative
products on a vector space $A$ (where $*$ may be partially defined, only)
and $\Delta $ a coassociative coproduct on $A$ is given if both $(A,*,\Delta
)$ and $(A,\cdot ,\Delta )$ are bialgebras and the following compatibility
condition between the products is satisfied for arbitrary elements $%
a,b,c,d\in A$: 
\[
(a*b)\cdot (c*d)=(a\cdot c)*(b\cdot d)
\]
whenever both sides are defined.
\end{definition}

\bigskip

Trialgebras were first suggested in \cite{CF} as an algebraic means for the
construction of four dimensional topological quantum field theories. It was
observed there that the representation categories of trialgebras have the
structure of so called Hopf algebra categories (see \cite{CF}) and it was
later shown explicitly in \cite{CKS} that from the data of a Hopf category
one can, indeed, construct a four dimensional topological quantum field
theory. The first explicit examples of trialgebras were constructed in \cite
{GS1} and \cite{GS2} by applying deformation theory, once again, to the
function algebra on the Manin plane and some of the classical examples of
quantum algebras and function algebras on quantum groups. In \cite{GS3} it
was shown that one of the trialgebras constructed in this way appears as a
symmetry of a two dimensional spin system. Besides this, the same trialgebra
can also be found as a symmetry of a certain system of infinitely many
coupled $q$-deformed harmonic oscillators.

We then have the following lemma:

\bigskip

\begin{lemma}
The data $\left( \mathcal{C},H\right) $ of \cite{KL} define for each choice
of $\mathcal{C}$ and $H$ a trialgebra.
\end{lemma}

%TCIMACRO{\TeXButton{Proof}{\proof}}
%BeginExpansion
\proof%
%EndExpansion
By definition of $\mathcal{C}$, $\mathcal{C}$ is equivalent to a category of
representations of an algebra $A$ (see \cite{KL}). So, on each object of $%
\mathcal{C}$ there is a representation $\cdot $ of the product of $A$. Now,
the Hopf algebra object $H$ is an object in $\mathcal{C}$ together with a
product $*$ and a coproduct $\Delta $. So, we have two associative products
and a coassociative coproduct given. It remains to check compatibilities.
The compatibility of $\Delta $ and $*$ is given by the definition of a Hopf
algebra object. The compatibility of $\Delta $ with $\cdot $ follows from
the fact that $\Delta $ is a morphism in $\mathcal{C}$. Finally, the
compatibility of $\cdot $ and $*$ follows also from the fact that $*$ is a
morphism in $\mathcal{C}$ together with the canonical extension of $\cdot $
to the tensor product of a representation of $A$ with itself.%
%TCIMACRO{\TeXButton{End Proof}{\endproof}}
%BeginExpansion
\endproof%
%EndExpansion

\bigskip

\begin{remark}
It was already observed in \cite{CF} that trialgebras define certain
monoidal bicategories. While two dimensional conformal field theories
without boundary arise from modular categories, the above lemma shows that
the two dimensional boundary conformal field theories which can be defined
through the construction of \cite{KL} correspond to a special class of
monoidal bicategories.
\end{remark}

\bigskip

As the next step, we can - in analogy to \cite{Dri} - ask for the universal
symmetry of (quasi-) trialgebras (where we allow for the coproduct and one
of the products to be quasi-associative, only) which are quasi-triangular
(respectively, coquasi-triangular) with respect to $\Delta $ and one of the
products. This question was considered in \cite{Sch} and it was found that
instead of $GT$ we get a Hopf algebra $\mathcal{H}_{GT}$ which was shown to
be self-dual, noncommutative, and noncocommutative. In addition, $\mathcal{H}%
_{GT}$ can be shown to be a sub-Hopf algebra of the Drinfeld double of $GT$.

Together with the above lemma, this shows that the action of the classical
group $GT$ is too restrictive to hold for general two dimensional boundary
conformal field theories. Instead, we have a quantum symmetry in the form of
an action of the Hopf algebra $\mathcal{H}_{GT}$, here. In this sense, the $%
GT$-symmetry on extended moduli space can not be physical since it does not
extend to general nonperturbative vacua. The $GT$-symmetry is extended in
this case to a quantum symmetry $\mathcal{H}_{GT}$.

\bigskip

\begin{remark}
Since the associator can be seen to basically determine the structure of $GT$
already (see \cite{Dri}, \cite{Kon 1999}), the assumption of
quasi-triangularity and coquasi-triangularity for the trialgebra which we
made above is not a decisive restriction.
\end{remark}

\bigskip

We conclude this section by a remark on some additional structural
properties of trialgebras: Though examples of trialgebras can be constructed
by further deformation quantization of quantum groups and the concept
carries some analogy to Hopf algebras (e.g dual pairings of trialgebras can
be introduced and a system of coupled matrix equations can be given,
replacing the $RTT$-relations in the trialgebraic case, see \cite{GS2}),
trialgebras are in several respects quite different objects.

To give a simple example of this kind, we mention that nontrivial
trialgebras are never unital. By a unital trialgebra one would mean one with
a unit element $1$ which is compatible with both products where the
algebraic notion of compatibility, usually used in the case of two products,
is the requirement that one and the same element $1$ should act as the unit
for both products. Such unital trialgebras are always trivial in the sense
that they are commutative bialgebras, i.e. the two products necessarily
agree and the product is commutative. The proof consists of a simple
Eckmann-Hilton type argument, for if such an element $1$ would exist, we
would have 
\begin{eqnarray*}
a\cdot b &=&\left( a*1\right) \cdot \left( 1*b\right) \\
&=&\left( a\cdot 1\right) *\left( 1\cdot b\right) \\
&=&a*b \\
&=&\left( 1*a\right) \cdot \left( b*1\right) \\
&=&b*a
\end{eqnarray*}
A much more involved result, showing that trialgebras are objects with new
and interesting algebraic properties, is given by the fact that - loosely
speaking - trialgebras can not be further deformed to algebraic structures
with e.g. two asscociative products and two coassociative coproducts, all
linked in a compatible way (see \cite{Sch}). In this sense, trialgebras are
the end of the story in the deformation process leading from groups to Hopf
algebras to trialgebras. We call this property of trialgebras \textit{%
ultrarigidity}.

\bigskip

\section{Cohomology of gerbes}

Geometrically, the introduction of boundary conditions in two dimensional
conformal field theory is related to the appearance of a 2-form field $B$
with a closed 3-form $H$ as its field strength. While closed 2-forms lead to
connections on vector bundles, closed 3-forms are interpreted as a kind of
connection on a gerbe (i.e. a kind of principal bundle with categories as
fibers). In \cite{BM} a detailed development of the necessary geometrical
theory of gerbes is started.

We will in this section be rather brief. Our aim is not to introduce the
heavy machinery of gerbes but mainly to show that the results, which have
been given in a completely rigorous way from an algebraic approach in the
foregoing section, can also be seen in this more geometric picture.

Following \cite{BM}, we consider a gerbe $\mathcal{P}$ over a base manifold $%
M$ with cover $\left( U_i\right) _{i\in I}$. Denote by $\mathcal{P}_{U_i}$
the fiber categories of $\mathcal{P}$ and let 
\[
U_{ij}=U_i\cap U_j 
\]
Then sections of $\mathcal{P}$ can be written as pairs of data $\left(
x_i,\phi _{ij}\right) $ with $x_i$ an object in $\mathcal{P}_{U_i}$ and 
\[
\phi _{ij}:x_j|_{U_{ij}}\rightarrow x_i|_{U_{ij}} 
\]
a morphism in $\mathcal{P}_{U_{ij}}$.

Let us now assume that $\mathcal{P}$ is a gerbe with gauge group (see \cite
{BM} for this concept) $GL\left( n,\Bbb{C}\right) $ (or $U\left( n\right) $
for the unitary case). Then we have the following result:

\bigskip

\begin{lemma}
For $\mathcal{P}$ a gerbe with gauge group $GL\left( n,\Bbb{C}\right) $ (or $%
U\left( n\right) $) over a smooth base manifold $M$, the sections of $%
\mathcal{P}$ define the structure of a $\mathcal{C}^\infty \left( M\right) $%
-linear ring category.
\end{lemma}

%TCIMACRO{\TeXButton{Proof}{\proof}}
%BeginExpansion
\proof%
%EndExpansion
$\mathcal{P}$ leads to a vector bundle analog $\widetilde{\mathcal{P}}$ of a
gerbe with sections $\left( \widetilde{x}_i,\widetilde{\phi }_{ij}\right) $
where the pointwise restrictions of $x_i$ and $\phi _{ij}$ lead to vector
spaces and linear maps, respectively. For a ring category we need a tensor
product $\otimes $ and a direct sum $\oplus $ with the obvious properties.
But these are induced from $\oplus $ and $\otimes $ of the category of $\Bbb{%
C}$-vector spaces. So, the category $\mathcal{S}$ of sections is a ring
category. $\Bbb{C}$-linearity of $\mathcal{S}$ follows from the fact that
pointwise we have $\Bbb{C}$-vector spaces and $\Bbb{C}$-linear maps. The
stronger result of $\mathcal{C}^\infty \left( M\right) $-linearity of $%
\mathcal{S}$ is a consequence of the fact that we have smooth data over $M$.%
%TCIMACRO{\TeXButton{End Proof}{\endproof}}
%BeginExpansion
\endproof%
%EndExpansion

\bigskip

\begin{remark}
This result is the analog of the fact that - by Serre duality - vector
bundles induce finite dimensional projective modules over $\mathcal{C}%
^\infty \left( M\right) $.
\end{remark}

\bigskip

Since there is a highly developed theory of Hochschild cohomology of
projective modules (see \cite{GeSch}), we can now ask for the analog of
Hochschild cohomology for such $\mathcal{C}^\infty \left( M\right) $-linear
ring categories $\mathcal{S}$. We will restrict to a simple but instructive
example, here (and make a remark about the more general case, below).
Namely, consider the case of such a category, consisting only of one object.
Since the morphism classes of $\mathcal{S}$ are $\Bbb{C}$-linear,
especially, it is immediately clear that $\mathcal{S}$ can be reinterpreted
as a $\Bbb{C}$-algebra $A$ where the product $\cdot $ of $A$ is just the
composition of $\mathcal{S}$. The tensor product gives a second product $*$
on $A$ and, using well known coherence theorems, we can assume without loss
of generality that $*$ is associative, too. We will call such a structure $%
\left( A,\cdot ,*\right) $ a double algebra, in the sequel. Observe that the
compatibility of $\cdot $ and $*$ is, once again, 
\begin{equation}
\left( a\cdot b\right) *\left( c\cdot d\right) =\left( a*c\right) \cdot
\left( b*d\right)  \label{1}
\end{equation}
for $a,b,c,d\in A$.

\bigskip

\begin{remark}
Since our aim is to consider deformation theory, in the sequel, we forget
about the additional sum $\oplus $ for the moment because the additive
structures remain fixed, anyway.
\end{remark}

\bigskip

Let us now come to the question of the cohomology of such a double algebra.
It is clear that $\left( A,\cdot ,*\right) $ defines two Hochschild
complexes - one for each product - but these are not independent but linked
by compatibility conditions for the cohomology groups induced from condition
(\ref{1}). E.g. for the second cohomology of $\left( A,\cdot ,*\right) $ we
do not have arbitrary pairs $\left( B_1,B_2\right) $ where $B_1$ is a
Hochschild 2-cocycle for $\cdot $ and $B_2$ a Hochschild 2-cocycle for $*$
but only pairs where $B_1$and $B_2$ satisfy the constraint 
\begin{eqnarray*}
&&\left( a\cdot b\right) *B_1\left( c,d\right) +B_1\left( a,b\right) *\left(
c\cdot d\right) -B_1\left( a*c,b*d\right) \\
&=&\left( a*c\right) \cdot B_2\left( b,d\right) +B_2\left( a,c\right) \cdot
\left( b*d\right) -B_2\left( a\cdot b,c\cdot d\right)
\end{eqnarray*}
calculated from the first order perturbation theory of condition (\ref{1}).

\bigskip

\begin{remark}
Actually, the structure of $\mathcal{S}$ induces for $A$ even the structure
of a double algebra over $\mathcal{C}^\infty \left( M\right) $ (instead of
simply a $\Bbb{C}$-linear double algebra). One could therefore suppose at
first sight that one can actually introduce, in addition, deformations of
the product in $\mathcal{C}^\infty \left( M\right) $ and that therefore
three different Hochschild complexes are involved. But observe that in the
undeformed case - by the composition of the linear maps $\widetilde{\phi }%
_{ij}$ - the composition in $\mathcal{S}$ and the product on $\mathcal{C}%
^\infty \left( M\right) $ are not independent but the product of $\mathcal{C}%
^\infty \left( M\right) $ can be seen as induced from the composition which
we would have for the case of $1\times 1$ matrices. In this sense, we do not
consider the product on $\mathcal{C}^\infty \left( M\right) $ as another
possibility for deformations and therefore restrict in the cohomology theory
to cohomology of a $\Bbb{C}$-linear double algebra.
\end{remark}

\bigskip

Next, remember that on Hochschild cohomology of associative algebras there
is - by the Deligne conjecture - a hidden action of the
Grothendieck-Teichm\"{u}ller group $GT$ (see \cite{Kon 1999} where a proof
of the Deligne conjecture is announced). Basically, the action of $GT$ can
be imagined as deriving from the possibility to weaken the associative
product to a quasi-associative one and from the transformation possibilities
for such an associator. We can therefore give the following heuristic
argument for the case of two associative products $\cdot $ and $*$: Here, we
have the possibility to introduce two different associators $\alpha ,\beta $
and the compatibility condition for $\alpha $ and $\beta $ is 
\begin{equation}
\left[ \alpha ,\beta \right] =0  \label{2}
\end{equation}
(observe that $\alpha $ and $\beta $ operate on the same space, i.e. we can
introduce a commutator for the successive application of the two maps). The
constraint resulting from condition (\ref{2}) for transformations of $\alpha 
$ and $\beta $ was calculated in \cite{Sch} and it was derived there that,
again, instead of $GT$ the Hopf algebra $\mathcal{H}_{GT}$ is the correct
algebraic object to describe the common transformations of $\alpha $ and $%
\beta $ respecting condition (\ref{2}).

Our heuristic argument therefore leads to the view that on the total
cohomology of a $\Bbb{C}$-linear double algebra $\left( A,\cdot ,*\right) $
we have to expect a hidden action of the Hopf algebra $\mathcal{H}_{GT}$.

\bigskip

\begin{remark}
We expect that the $\mathcal{H}_{GT}$-action should also hold for the case
of a general category with the structure of $\mathcal{S}$ because in the
case of the $GT$-action this is also not affected by the passage from
associative algebras to $A_\infty $-categories.
\end{remark}

\bigskip

The main conclusion of this section is therefore that also in the more
geometric picture of gerbes we find an argument that the $GT$-symmetry can
not be physical in the sense that it does not hold for general
nonperturbative backgrounds but that it has to be extended to the quantum
symmetry $\mathcal{H}_{GT}$ in this case. The arguments of this section are
of a more heuristic nature but one should keep in mind that we have given a
completely rigorous approach in the algebraic framework in the previous
section. The aim of this one was mainly to convey the general view which
presents itself from the geometrical side, in this question.

Let us conclude this section by passing to the question if the $\mathcal{H}%
_{GT}$-symmetry is stable against quantization. In this part, we can only
give a heuristic argument at this time. We will argue that for the $\mathcal{%
H}_{GT}$-symmetry we should not only have stability against quantization but
that in this case the quantum and the classical description should even be
equivalent.

The argument relies on the property of ultrarigidity introduced in \cite{Sch}%
. By ultrarigidity we can not expect a $\mathcal{H}_{GT}$-invariant theory
to have a nontrivial quantum deformation. Since any such deformation should
be equivalent to the nondeformed theory, we expect that for a $\mathcal{H}%
_{GT}$-invariant theory the quantum theory should even be equivalent to the
classical one.

\bigskip

\section{Conclusion}

We have in this paper given arguments from two different perspectives - the
algebraic formulation of boundary conformal field theories of \cite{KL} and
the cohomology of gerbes - that the $GT$-symmetry on extended moduli space
can not represent a physically relevant symmetry but that upon inclusion of
general nonperturbative vacua it has to be extended to a quantum symmetry
represented by a self-dual, noncommutative, and noncocommutative Hopf
algebra $\mathcal{H}_{GT}$.

A more detailed investigation of the implications of a universal $\mathcal{H}%
_{GT}$-symmetry in open string theory will follow in subsequent work.

\bigskip

\textbf{Acknowledgements:}

I thank the Erwin Schr\"{o}dinger Institute for Mathematical Physics,
Vienna, for hospitality during the time where this work has been done. For
discussions on the subjects involved, I thank Bojko Bakalov, Karen Elsner,
J\"{u}rgen Fuchs, Harald Grosse, Christoph Schweigert, and Ivan Todorov.

\end{document}